\documentclass[twocolumn]{aastex631}
\submitjournal{ApJ}
\usepackage[T1]{fontenc}
\usepackage{ae,aecompl}
\usepackage{graphicx}
\usepackage{amsmath}	
\usepackage{amssymb}	
\usepackage{tikz}
\usepackage{booktabs}
\usepackage{rotating}

\begin{document}
\label{firstpage}
\title{Automated Detection of Galactic Rings from SDSS Images}

\author{Linn Abraham}
\affiliation{Inter-University Centre for Astronomy and Astrophysics, IUCAA Pune, India}
\affiliation{Puducherry Technological University, Puducherry, India}
\email{linn.official@gmail.com}

\author{Sheelu Abraham}
\affiliation{Inter-University Centre for Astronomy and Astrophysics, IUCAA Pune, India}
\affiliation{Marthoma College, Chungathara, Nilambur, Kerala, India}
\email{sheeluabraham@mtcc.ac.in}

\author{Ajit K. Kembhavi}
\affiliation{Inter-University Centre for Astronomy and Astrophysics, IUCAA Pune, India}
\email{akk@iucaa.in}

\author{N. S. Philip}
\affiliation{Inter-University Centre for Astronomy and Astrophysics, IUCAA Pune, India}
\affiliation{Artificial Intelligence Research and Intelligent Systems, Kerala, India}
\email{ninansajeethphilip@airis4d.com}

\author{A. K. Aniyan}
\affiliation{DeepAlert Ltd, Bromley, BR1 1QE, London, UK}
\email{arun@deepalert.ai}

\author{Sudhanshu Barway}
\affiliation{Indian Institute of Astrophysics, Bengaluru, India}
\email{sudhanshu.barway@iiap.res.in}

\author{Harish Kumar}
\affiliation{Puducherry Technological University, Puducherry, India}
\email{harishkumarholla@ptuniv.edu.in}

\begin{abstract}
Morphological features in galaxies, like spiral arms, bars, rings, tidal tails etc. carry information about their structure, origin and evolution.  It is therefore important to catalog  and study  such  features  and to correlate them with other basic galaxy properties, the environment in which the galaxies are located and their interactions with other galaxies.
The volume of present and future data on galaxies is so large that traditional methods, which involve expert astronomers identifying morphological features through visual inspection, are no longer sufficient. It is therefore necessary to use AI based techniques like machine learning and deep learning for finding morphological structures quickly and efficiently.  We report in this study the application of deep learning for finding ring like structures in galaxy images from the Sloan Digital Sky Survey (SDSS) data release DR18.  We use a catalog by Buta (2017) of ringed galaxies from the SDSS
to train the network, reaching good accuracy and recall, and generate a catalog of 29420 galaxies of which 4855 have ring like structures with prediction confidence exceeding 90 percent.  Using a catalog  of barred galaxy images  identified  by Abraham et. al. (2018) using deep learning techniques, we identify a set of 2087 galaxies with bars as well as rings.  The catalog should be very useful in understanding the origin of these
important morphological structures.  As an example of the usefulness of the catalog, we  explore the environments and 
star formation characteristics of ring galaxies in our sample.
\end{abstract}


\section{Introduction}
Galaxies display a wide variety of forms as a consequence of the differences in their intrinsic structure, interaction with other galaxies as well as observational biases.
Efforts to make sense of galaxy morphology with its multitude of forms
have a long history dating back to Edwin Hubble and others.
The Hubble classification system \citep{1926ApJ....64..321H} arranged galaxies in a sequence ranging from ellipticals to lenticulars to spirals to irregulars shaped objects.
Galaxies other than the ellipticals were further classified  into those with bar like structures and those without such structures, resulting in a  tuning fork like diagram.  Other more detailed classification systems have been developed \citep[see, e.g.,][]{van_den_bergh_galaxy_1998}, taking into account further morphological features.

In recent decades, the availability of many millions of higher resolution multiwavelength images from large deep field surveys with modern instrumentation have made possible detailed morphological studies \citep{buta_galaxy_2011}.
Galaxy morphology plays a key role in bettering our understanding of the secular processes that underlie galaxy evolution.  Several important questions concerning the formation and evolution of galaxies can be addressed from  the observed change in morphology with epoch.

A fairly common feature of disk galaxies is the presence of a ring shaped pattern in their light distribution \citep{buta_galactic_1996}.
These rings in galaxies have come to be considered an integral part of galaxy morphology \citep{kormendy_morphological_1979}. Recognizing their importance, the de Vaucouleurs Revised Hubble-Sandage Classification System \citep[VRHS;][]{flugge_classification_1959} 
added rings as another dimension to the two-dimensional Hubble tuning fork, turning it into a classification volume. Rings in galaxies can fundamentally be divided into normal rings, also called `resonance rings' and `catastrophic rings', which are a result of galaxy collisions \citep{buta_galactic_1996}. Although different sub-classes exist within the catastrophic rings, such as polar, accretion, and collisional rings, they constitute a very small minority of all observed rings. \citet{2009yCat..21810572M} have estimated the abundance of collisional rings to be only about 1 in 1000. Several theories have been put forward to explain the different ring structures that are observed, with the resonance interpretation being the most popular. The ``Manifold Theory'' that has been proposed  \citet{10.1111/j.1365-2966.2008.14273.x} also has had considerable success in explaining several aspects of ring morphology.

Several catalogs that contain the visual morphology classification of galaxies are available today. Notable among them are
\citet{nair_catalog_2010} and 	\cite{Fukugita_2007}, with the former containing the detailed classifications of 14,034 galaxies.
Crowd-sourcing efforts such as the Galaxy Zoo citizen science project \citep{lintott_galaxy_2011} have managed to ramp up the number of galaxy morphology classifications to nearly 900,000 by involving volunteers. This vast database has motivated others to pick specific subtypes of galaxies for detailed classifications. For example, \citet{buta_galactic_2017-1} has exhaustively classified 3962 ringed galaxies taken from this set.
However, all these efforts involve a human in the loop for classifying galaxy morphologies.

With an explosion  in the number of galaxy images being produced as part of imaging surveys like the Sloan Digital Sky Survey \citep[SDSS;][]{york_sloan_2000}, Panoramic Survey Telescope and Rapid Response System \citep[Pan-STARRS;][]{2002SPIE.4836..154K}, Dark Energy Survey \citep[DES;][]{dark_energy_survey_collaboration_dark_2016} etc.,  automated methods of morphology classification have become very important. Although traditional image processing techniques, many of which are histogram-based, have had some success \citep[see, e.g.,][] {shamir_automatic_2020}, these are inappropriate for complex tasks such as shape recognition from images. However, the advent of artificial intelligence (AI) techniques like Convolutional Neural Networks \citep[CNNs;][]{lecun1995convolutional} has significantly improved the automated efforts for galaxy morphology classification.
One of the initial efforts in this area have been made by \citet{banerji_galaxy_2010}.
\citet{dieleman_rotation-invariant_2015} have used CNNs to create a model that can match the combined consensus of the Galazy Zoo volunteers on all of the questions regarding galaxy morphology that are part of the survey.
\cite{abraham_detection_2018,sanchez_improving_2018} have used CNNs for the automated classification of galaxy morphological features such as bars, bulges, edge-on morphology etc.
More recently, \cite{shimakawa_galaxy_2023} have used a deep-learning model to obtain ring classification for a very large sample of galaxies taken from the Hyper Suprime-Cam Subaru Strategic Program.

In the present study we investigate the efficacy of an automated method based on Convolutional Neural Networks for the detection of rings in galaxies.
We use galaxy images from the Sloan Digital Sky Survey \citep[SDSS;][]{york_sloan_2000} for the training and testing of our network.
We treat the problem as a binary classification one,  treating  all sub-classes of rings as a single {\tt ring} type and distinguishing it from galaxies that do not contain any rings, the {\tt non-ring} type.  The trained network can be used to identify ringed galaxies from the SDSS.
We used the network's predictions to generate a sample of 4855 ringed galaxies with a classification threshold of 0.90.
Using this sample we explored the connection between ring formation and star formation activity considering the main sequence, green valley and quenched galaxy populations.
We also analyzed the star formation by examining their distribution across different environments, categorized by their local surface densities.

The rest of the paper is organized as follows:
In Section \ref{data} we describe the data used in the analysis   and introduce the 
neural network architecture in Section \ref{network}.
We describe the  data augmentation techniques and   training procedure for the neural network in Sections \ref{augmentation} and \ref{training} respectively and in 
Section \ref{results} we analyse the  results on the trained network. 
In Section \ref{catalogue_rings} we present the catalog of ringed galaxies generated using our trained network and in 
Section \ref{catalogue_barred_rings} we present a subset of the ringed galaxies that have also been identified to have bars.
In  Section \ref{discussion} we provide a discussion regarding the environments and star formation rates for the ringed galaxies in our catalog.
Finally in Section \ref{conclusion} we give the concluding remarks.

\section{Data} \label{data}
The performance of a supervised machine learning classifier depends to a large extent on the quality of the labelled data that is made available to it. For deep learning models the quantity becomes important as well because of the large number of trainable or free parameters in the model.
Many of the standard datasets available for galaxy morphology classification  do not contain rings as a class
(see, e.g., DECALS10 \footnote{\url{https://astro.utoronto.ca/~hleung/shared/Galaxy10/Galaxy10_DECals.h5}}).
Even with the Galaxy Zoo 2 dataset \citep{willett_galaxy_2013}, ringed galaxies for which at least 50 percent of zoo volunteers agreed on a response are relatively small in number.

\subsection{Rings and Non-Rings}
We used the \citet{buta_galactic_2017-1} catalog as our primary source to identify galaxies with rings.
The author has provided detailed visual morphological classification of 3962 galaxies such galaxies.
The classifications are done within the framework of the comprehensive de Vaucouleurs revised Hubble-Sandage system \citep[CVRHS;][]{buta_classical_2015, buta_vaucouleurs_2007}.
The catalog also provides the author's comments that are useful for creating a good quality training set by identifying rare cases of rings, poorly resolved galaxies etc.
The galaxies used for classification by the author were picked from the Galaxy Zoo 2 \citep{willett_galaxy_2013} survey (hereafter, GZ2).

Along with a good quality training sample for rings, it is equally important to obtain a good sample of galaxies that do not have any presence of rings.
Since the \citet{buta_galactic_2017-1} catalog does not contain such galaxies, we used the \citet{nair_catalog_2010} catalog, which is one of the largest catalogs for the visual morphological classification of galaxies.
They used galaxy images from the SDSS DR4 release (\cite{stoughton_sloan_2002}; \cite{adelman-mccarthy_fourth_2006}).
Galaxies with spectroscopic redshifts in the range $0.01 < z < 0.1$ and extinction corrected \emph{g'}-band magnitude  $<16$ were selected from the spectroscopic main sample \citep{strauss_spectroscopic_2002}.
This led to their final sample size of 14,034 galaxies.

\begin{figure*}
    \centering
        \includegraphics[width=\textwidth]{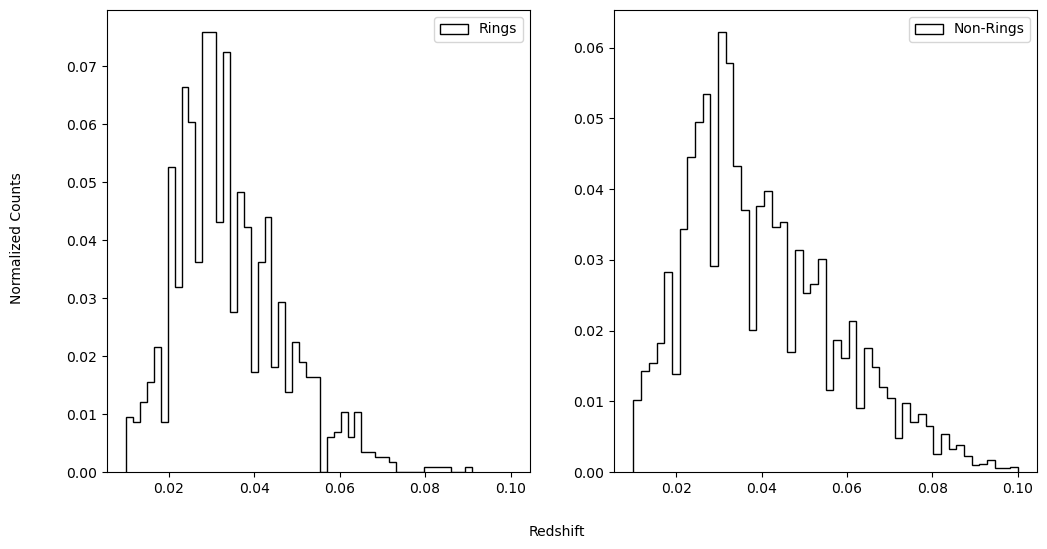}
        \label{fig:buta_z_new}
        \includegraphics[width=\textwidth]{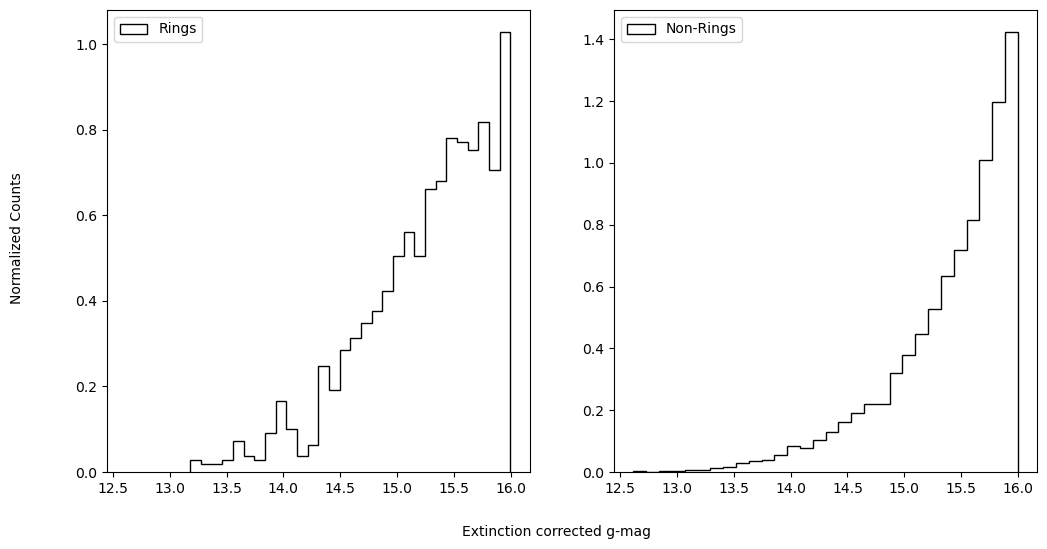}
        \label{fig:gmag_dist_new}
    \caption{The normalized distribution of redshifts and extinction corrected g-magnitude for the ringed galaxies(left) and non-ringed galaxies(right)}
    \label{fig:distribution}

\end{figure*}

The classification scheme used
in \citet{nair_catalog_2010} is not the CVRHS. Rather it is primarily based on the Carnegie Atlas of Galaxies
\citep{1994cag..book.....S}
used in consultation with the Third Reference Catalog of Bright Galaxies
\citep[RC3;][]{1991rc3..book.....D}.
In order to use only galaxies without rings, we selected from the catalog galaxies that have the ring type and ring flag column set to zero.
\subsection{Selection Criteria}
The images used for the GZ2 survey come from the SDSS Data Release 7 \citep[DR7;][]{abazajian_seventh_2009}.
The GZ2 includes galaxies with extinction corrected Petrosian half-light magnitude in the r-band < $17.0$, Petrosian radius $\tt{petroR90_r}>3$ arcsec and spectroscopic redshift in the range $0.0005 < z < 0.25$, when it is known.
Finally, galaxies that have SDSS flags which are either SATURATED, BRIGHT, OR BLENDED without an accompanying NODEBLEND were removed.
We further selected galaxies that have an extinction corrected \emph{g'}-band magnitude < 16 based on the selection criteria used for the non-ring galaxies.
The spectroscopic redshift was also limited to be in the same range as that of the non-ring galaxies, specifically $0.01 < z < 0.1 $.
By visual inspection we removed 37 ringed and 797 non-ringed galaxies which contained artefacts and those which were not clearly distinguishable.
After the selection criteria was applied and the manual removal, the number of ringed galaxies obtained were 1122 and the number of galaxies without rings came out to be 10,639.
Figure \ref{fig:distribution} shows the distribution of redshift and extinction corrected g-magnitude for the ringed and non-ringed galaxies in our training set.

\subsection{SDSS Image Cutouts}

\begin{figure*}
    \begin{minipage}{0.98\textwidth}
        \centering
        \includegraphics[width=\textwidth]{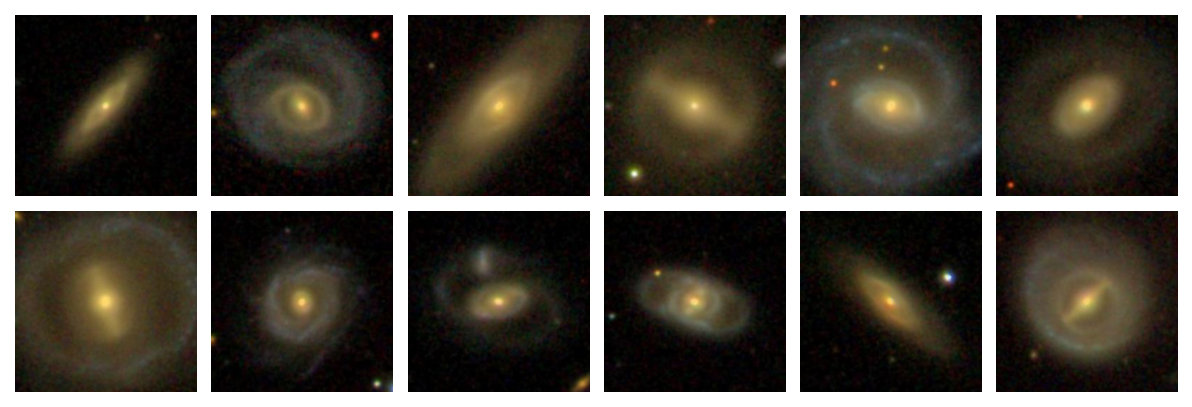}
        Galaxies with rings
        \label{fig:r-sample}
    \end{minipage}
    \begin{minipage}{0.98\textwidth}
        \centering
        \includegraphics[width=\textwidth]{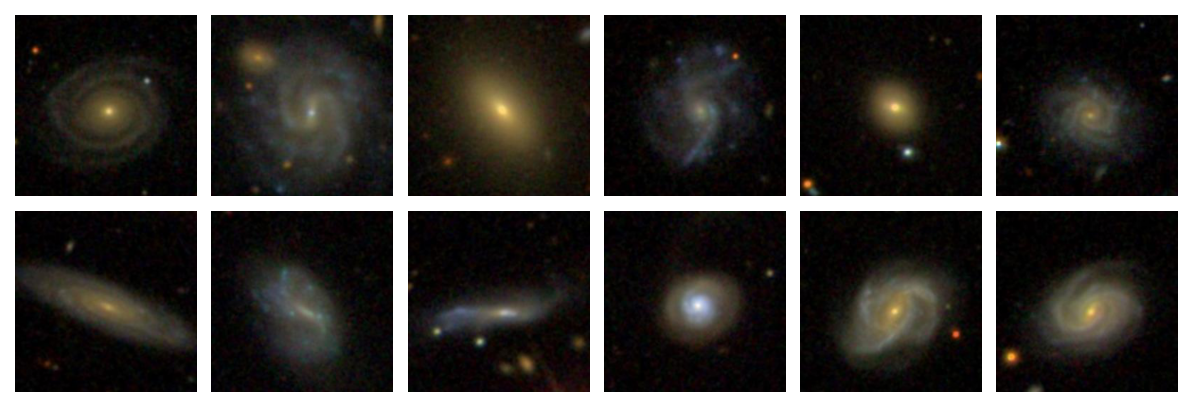}
        Galaxies with no rings
        \label{fig:nr-sample}
    \end{minipage}
    \caption{A sample of galaxies with and without rings from our training set.}
    \label{fig:galaxies_train}

\end{figure*}

 We modified a python library\footnote{\url{https://pypi.org/project/panstamps/}} that exists in the public domain for downloading images we needed for training as well as catalog preparation. We incorporated multi-threaded parallel downloading using the {\tt Concurrent} \footnote{\url{https://docs.python.org/3/library/concurrent.futures.html}} python module into the script to speed up the downloading process. 
Figure \ref{fig:galaxies_train} shows a sample of the galaxies with and without rings in our training set.

\section{Network Design} \label{network}

AlexNet \citep{krizhevsky_imagenet_2012} is one of the earliest CNN architectures designed and yet one of the most reliable. Having a smaller size than most modern architectures results in lesser training time and also the requirement of lesser amount of training data for this network.
The original AlexNet architecture was designed for the ImageNet challenge \citep{deng_imagenet_2009}.
The network has eight layers with the first five being convolutional layers and the remaining three being fully-connected.
The first convolutional layer filters the 224 × 224 × 3 input image with 96 kernels of size 11 × 11 × 3 with a stride of four pixels.
All of the convolution and fully connected layers are immediately followed by a block consisting of a Rectified Linear Unit activation (ReLU; \citealt{nair_rectified_2010}) and a batch normalisation (NORM) layer.
The first, second and final convolution layers are each followed by a maxpooling layer, that leads to 50 percent downsampling of the convolution feature maps.
These layers also contribute to the rotational invariance of the learnt model weights \citep{boureau_theoretical_2010}.
The max-pooling layers of the first two convolutions are followed by dropout layers \citep{srivastava_dropout_2014} having a dropout fraction of 0.25.
These prevent overfitting of the model by randomly dropping 25 percent of the weights during each training iteration.

\begin{figure*}
    \centering
    \begin{minipage}{0.3\linewidth}
    \centering
\begin{tabular}{|l|l|l|l|}
    \toprule
    {} &                   Name &                Type &                 Shape \\
    \midrule
    0  &                 conv2d &              Conv2D &  (None, 100, 100, 96) \\
    1  &             activation &          Activation &  (None, 100, 100, 96) \\
    2  &    batch\_normalization &  BatchNormalization &  (None, 100, 100, 96) \\
    3  &          max\_pooling2d &        MaxPooling2D &    (None, 49, 49, 96) \\
    4  &                dropout &             Dropout &    (None, 49, 49, 96) \\
    5  &               conv2d\_1 &              Conv2D &   (None, 49, 49, 256) \\
    6  &           activation\_1 &          Activation &   (None, 49, 49, 256) \\
    7  &  batch\_normalization\_1 &  BatchNormalization &   (None, 49, 49, 256) \\
    8  &        max\_pooling2d\_1 &        MaxPooling2D &   (None, 24, 24, 256) \\
    9  &              dropout\_1 &             Dropout &   (None, 24, 24, 256) \\
    10 &               conv2d\_2 &              Conv2D &   (None, 24, 24, 384) \\
    11 &           activation\_2 &          Activation &   (None, 24, 24, 384) \\
    12 &  batch\_normalization\_2 &  BatchNormalization &   (None, 24, 24, 384) \\
    13 &               conv2d\_3 &              Conv2D &   (None, 24, 24, 384) \\
    14 &           activation\_3 &          Activation &   (None, 24, 24, 384) \\
    15 &  batch\_normalization\_3 &  BatchNormalization &   (None, 24, 24, 384) \\
    16 &               conv2d\_4 &              Conv2D &   (None, 24, 24, 256) \\
    17 &           activation\_4 &          Activation &   (None, 24, 24, 256) \\
    18 &  batch\_normalization\_4 &  BatchNormalization &   (None, 24, 24, 256) \\
    19 &        max\_pooling2d\_2 &        MaxPooling2D &   (None, 11, 11, 256) \\
    20 &              dropout\_2 &             Dropout &   (None, 11, 11, 256) \\
    21 &                flatten &             Flatten &         (None, 30976) \\
    22 &                  dense &               Dense &          (None, 4096) \\
    23 &           activation\_5 &          Activation &          (None, 4096) \\
    24 &  batch\_normalization\_5 &  BatchNormalization &          (None, 4096) \\
    25 &              dropout\_3 &             Dropout &          (None, 4096) \\
    26 &                dense\_1 &               Dense &          (None, 4096) \\
    27 &           activation\_6 &          Activation &          (None, 4096) \\
    28 &  batch\_normalization\_6 &  BatchNormalization &          (None, 4096) \\
    29 &              dropout\_4 &             Dropout &          (None, 4096) \\
    30 &                dense\_2 &               Dense &             (None, 2) \\
    31 &           activation\_7 &          Activation &             (None, 2) \\
    \bottomrule
\end{tabular}
\end{minipage}
\begin{minipage}{0.9\linewidth}
\centering
    \includegraphics[width=\textwidth]{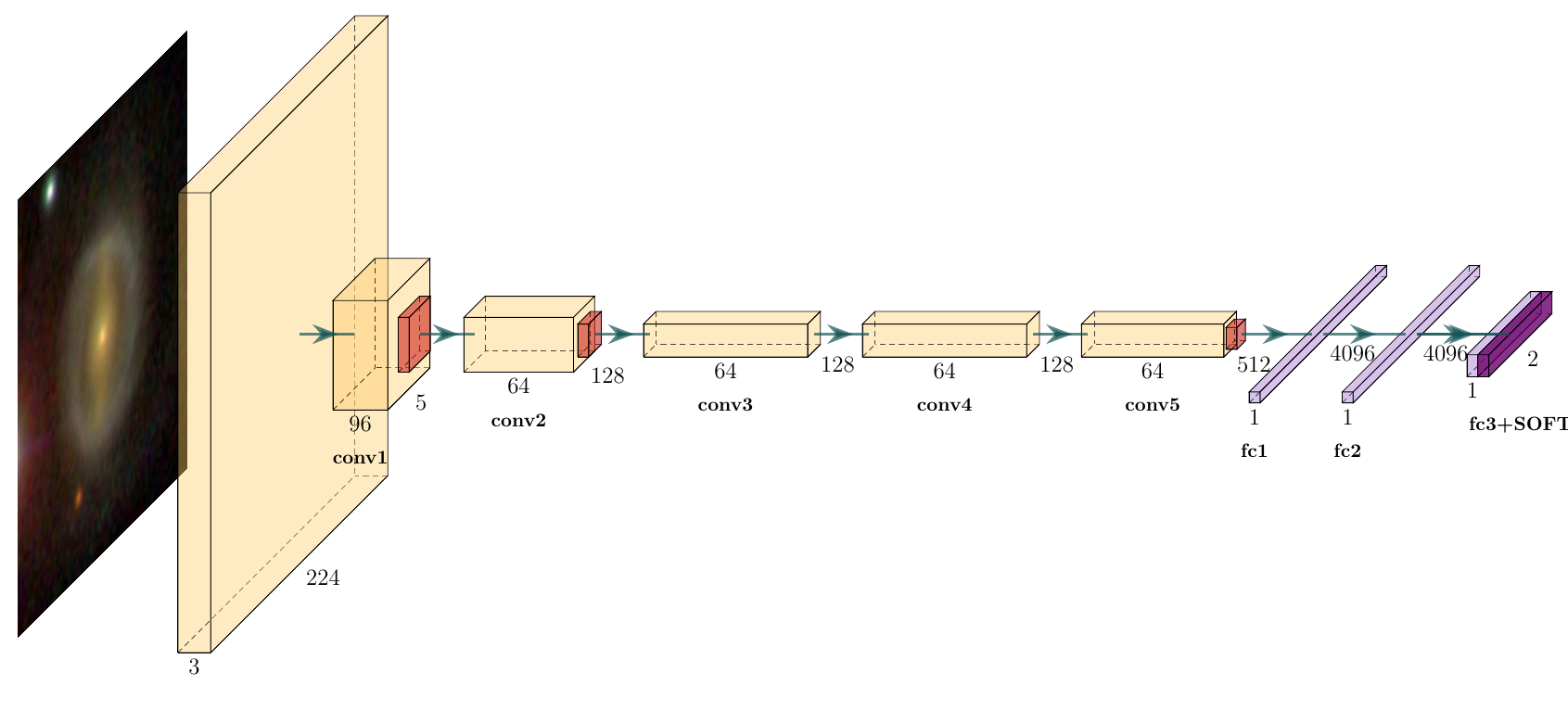}
\end{minipage}
\hfill
\caption{A tabular description of the AlexNet model architecture that we have used, generated with the {\tt Keras} package (above) and a visual representation of the same (below). Code for generating the figure was adapated from \protect \citet{iqbal_harisiqbal88plotneuralnet_2018} }
    \label{fig:model-arch}
\end{figure*}

We left the input layer size unchanged since it was suitable for our purposes.
However, galactic rings are features that are harder to distinguish in comparison to the classes in the ImageNet challenge.
We therefore used a smaller size of five for the kernels in the first convolutional layer. This seemed to have a considerable impact on the accuracy of the network.
Thus in our network the first convolution layer has 96 kernels of size $ 5 \times 5 \times 3 $ and the output is fed to the second layer of convolution with 256 kernels each of size $5 \times 5 \times 48$ .
The third, fourth and fifth convolutional layers  have respectively 384, 384 and 256 kernels of size $3 \times 3 \times 256$, $3 \times 3 \times 192$ and $3 \times 3 \times 192$ respectively.
These learn more complex features and their final output is fed into another max-pooling layer.
The output of the final pooling layer is then fed into a series of fully connected layers of size 4096.
These layers are connected by dropout layers which have 0.50 dropout fraction.
The final layer is the sigmoid activation function layer which outputs a value that can be interpreted as the probability of the input belonging to the positive class.
The architecture of our network is shown as a table and schematic diagram in Figure \ref{fig:model-arch}.

\subsection{Selection of Hyperparameters}

Hyperparameters are those parameters that are not learnt or updated during the learning process and hence need to be manually set by the programmer. 
These can include the network architecture itself and several other parameters like the batch size used for the stochastic gradient descent or the number of epochs of training etc.  Hyperparameter tuning is an important part of a deep learning project, since these parameters have a direct bearing on the performance and accuracy of the model.

The loss function, which  specifies the error and how much weightage should be given to individual errors is an important hyperparameter to tune.
We have used {\tt categorical crossentropy}\footnote{\url{https://www.tensorflow.org/api_docs/python/tf/keras/metrics/categorical_crossentropy}}, which is a standard loss function suitable for binary classification problems.
The optimizer function has an effect on how the network weights are updated in order to minimize this loss. We used the Adam optimizer with a learning rate of 0.003.
Regularization techniques are a set of safeguard measures that prevent the model from overfitting on the training data and thus losing its ability to generalize.
We have used L2 weight regularization with a factor of 0.0002 to penalize the weights from growing too much during training.
The {\tt batch size} controls how many input samples are considered together for evaluating the loss function. This leads to a trade-off between accuracy and speed.  A batch size of 16 was seen to be sufficient in our case. The number of epochs or ({\tt num epochs}) decides how many times the network gets to see the complete training data.  We trained the network for {\tt num epochs} without early stopping.

\section{Image augmentation} \label{augmentation}
\begin{figure*}
    \begin{minipage}{0.15\textwidth}
    \centering
        \includegraphics[width=\textwidth]{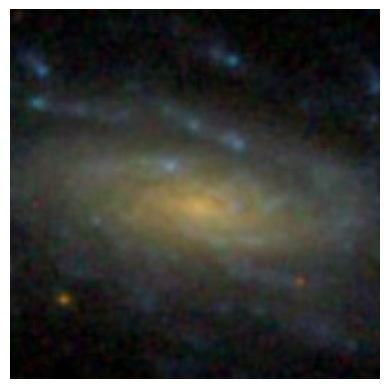}
        Original
        \label{fig:aug_org}
\end{minipage}
    \begin{minipage}{0.15\textwidth}
    \centering
        \includegraphics[width=\textwidth]{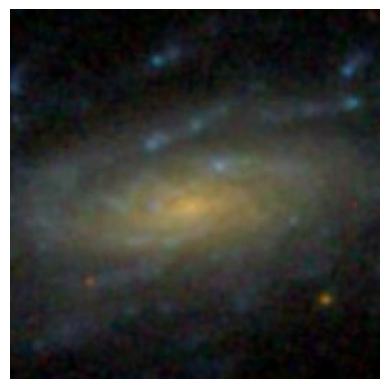}
        Horizontal Flip
        \label{fig:aug_hflip}
\end{minipage}
    \begin{minipage}{0.15\textwidth}
    \centering
        \includegraphics[width=\textwidth]{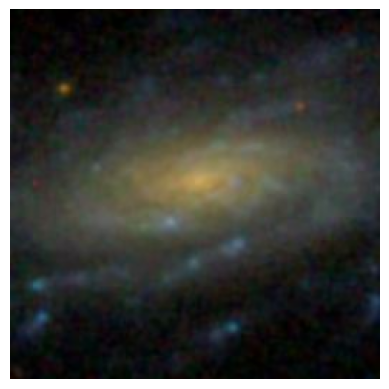}
        Vertical Flip
        \label{fig:aug_vflip}
\end{minipage}
    \begin{minipage}{0.15\textwidth}
    \centering
        \includegraphics[width=\textwidth]{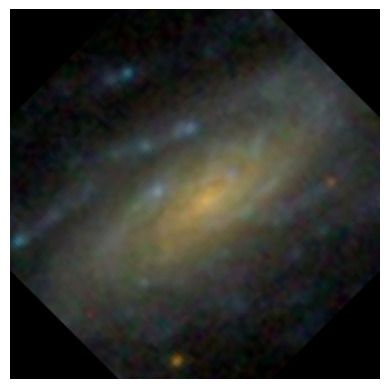}
        Rotation
        \label{fig:aug_rot}
\end{minipage}
    \begin{minipage}{0.15\textwidth}
    \centering
        \includegraphics[width=\textwidth]{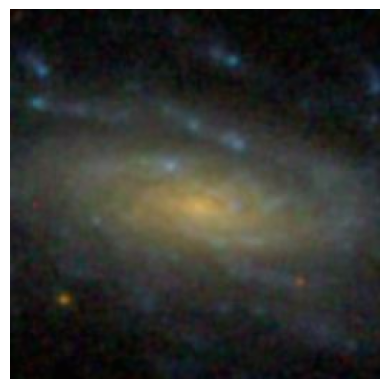}
        Brightness
        \label{fig:aug_bright}
\end{minipage}
    \begin{minipage}{0.15\textwidth}
    \centering
        \includegraphics[width=\textwidth]{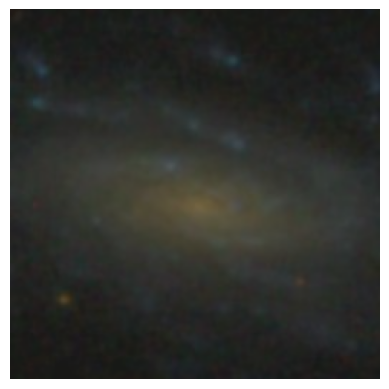}
        Contrast
        \label{fig:aug_cchange}
\end{minipage}
    \caption{The different types of image augmentations that are done to each image in our training set.}
    \label{fig:augment}

\end{figure*}

Neural networks are prone to overfitting if the data available for training is insufficient.  
In our case the possibility of increasing the training data is limited by the availability of  catalogs of ringed galaxies.
It is therefore necessary to   increase the training sample size, through data augmentation by transforming the available images.  We have used operations such as horizontal and vertical flipping, rotation through arbitrary angles, brightness and contrast adjustments for augmenting our training set. Using rotations also makes the network invariant to rotations, which is needed because of the rotational symmetry of the ringed galaxies,  The augmentations that we have used are illustrated in Figure \ref{fig:augment} using a random galaxy.

We implemented image augmentations using the {\tt on-the-fly} mode provided by the {\tt keras-tensorflow} \citep{chollet2015keras} framework.  Here a predefined sequence of transformations is applied to each image from the training set, using  parameters which are randomized so that at each epoch, the network sees a different set of augmented images.  In this way we avoid generating all the augmented images before the start of the training and also save disk space.

\section{Training} \label{training}

The galaxy image cutouts used for training are 3 channel RGB $256 \times 256$ pixel images which are 
are dynamically resized by the {\tt Keras}
{\tt image\_dataset\_from\_directory} \footnote{\url{https://www.tensorflow.org/api_docs/python/tf/keras/utils/image_dataset_from_directory}} function into the required input shape of $240 \times 240$. 
The pixel intensities in the range (0-255) are  rescaled to the range (0,1) for easier computation.
The input size and other parameters for the network like {\tt num epochs}, {\tt batch size} are controlled through a config file which is read at run time. The images are first resized to the desired input size and then loaded into memory in {\tt nbatches} using the Keras {\tt image\_dataset\_from\_directory} function. A generator avoids the need for all of the images to be loaded into memory at once.

\begin{figure*}
    \centering
    \includegraphics[width=0.45\textwidth]{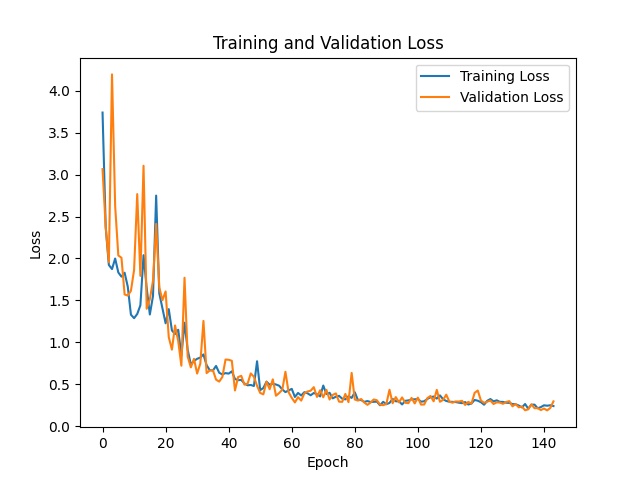}
    \includegraphics[width=0.45\textwidth]{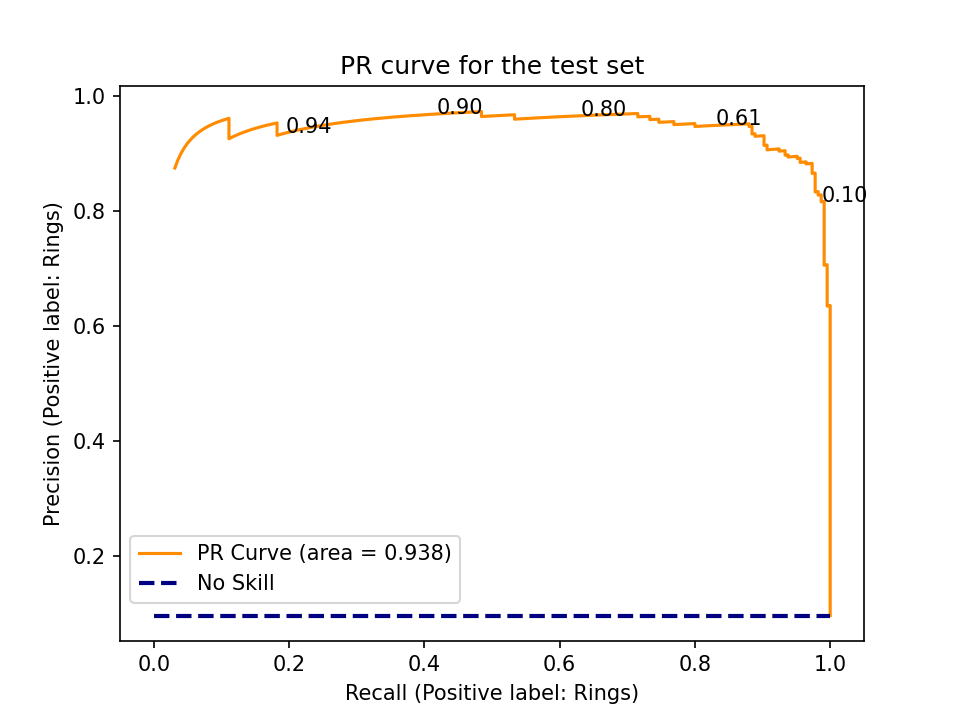}
     \begin{tabular}{|l|l|l|l|}
         \toprule
         {} Class &            Precision(\%)        &    Recall(\%)                   & F1-Score(\%) \\
         \midrule
         Rings  &                 93 &              89 &  91 \\
         Non-Rings &              99 &              99 &  99 \\
         \bottomrule
     \end{tabular}
    \caption{(1) The training graph, which shows a plot of the training and validation loss against the number of epochs (top-left). (2) The Precision-Recall curve which shows the plot of precision against recall for ringed galaxies in our test set computed at various thresholds (top-right). The dashed line shows the curve expected from a no-skill classifier. A few thresholds taken at random are annotated on the curve. (3) The classification report showing the precision and recall for both rings and non-rings in our test set (bottom).
    }
    \label{fig:train-results}
\end{figure*}

The data is initially split into a training and testing set using a ratio of 80:20.
During training the training set is further split internally into a training and validation set using the same ratio.
Before augmentations are applied, the number of ringed galaxies in our training set (including the validation split) is 897 and those in the testing set is 225.
Similarly, the number of galaxies without rings in our training set is 8511 and those in the testing set is 2128.
The network was trained for a total of 140 epochs.
The training graph in Figure \ref{fig:train-results} shows the variation of the {\tt categorical\_crossentropy} loss function with increasing epochs for both the training and validation data.  If there is overfitting, the validation loss of the network increases whereas the training loss keeps decreasing.
In our case, the training loss and validation loss are seen to decrease together with each epoch and therefore there is no significant overfitting that occurs.  There is further scope for training the network albeit with minimal returns.

The training was done on the Amazon Web Services platform on an Intel Xeon workstation with 4 virtual CPUs and 16 GB RAM, making use of a Tesla (g4dn.xlarge) GPU accelerator with 16 GB graphics memory.
A community AMI (Amazon Machine Image) running Arch Linux was modified to function as our base operating system.
The data and model are version controlled using Data Version Control  \cite[DVC;][]{ruslan_kuprieiev_2022_7275755} and stored on Amazon S3 cloud storage.
The code itself is version controlled and made available on github \citep{abraham_automated_2024-1}.
The entire training session takes 2.5 hours to run with our hardware specifications.

\section{Results} \label{results}
\subsection{Evaluation Metrics for Unbalanced Datasets}
For evaluating the performance of the network in discriminating between galaxies with rings and those without rings, we us the three metrics accuracy, precision and recall.
Accuracy, is the number of correct classifications as a fraction of the total number of classifications made.
$$ \rm{Accuracy} = \frac{TP+TN}{TP+TN+FP+FN}$$
where TP denotes the number of true positives, FP the number of false positives and FN the number of false negatives.
Precision and recall are defined as
$$ \rm{Precision} = \frac{TP}{TP + FP} $$
$$ \rm{Recall} = \frac{TP}{TP+FN}$$
Precision is an indicator of the purity of the classification, that is the extent to which a collection of galaxies identified as ringed are indeed ring galaxies.  Recall on the other hand is a measure of the extent of completeness in identifying galaxies from a sample containing galaxies with and without rings.  Precision and recall are therefore important for astronomers.
 
\begin{figure*}
    \centering
    \begin{minipage}{0.93\textwidth}
        \centering
        \includegraphics[width=0.96\textwidth]{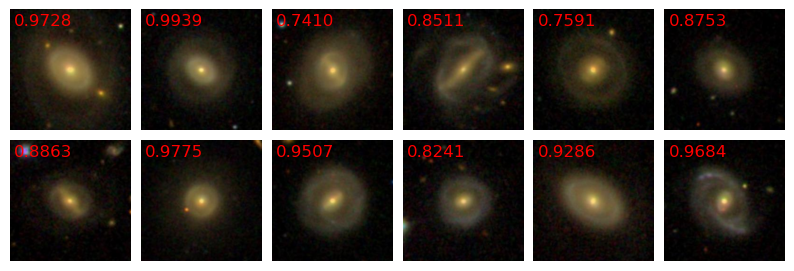}
        True positives
        \label{fig:tp}
    \end{minipage}
    
    \begin{minipage}{0.93\textwidth}
        \centering
        \includegraphics[width=0.96\textwidth]{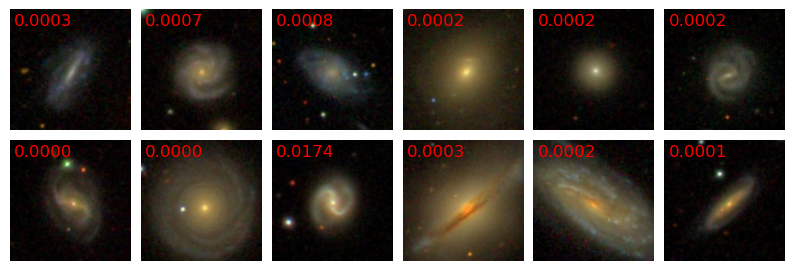}
        True negatives
        \label{fig:tn}
    \end{minipage}
    
    \begin{minipage}{0.93\textwidth}
        \centering
        \includegraphics[width=0.96\textwidth]{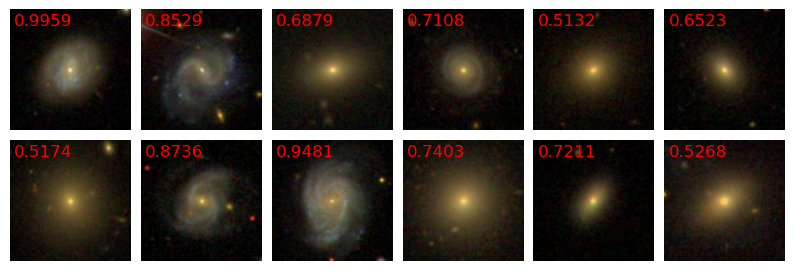}
        False positives
        \label{fig:fp}
    \end{minipage}

    \begin{minipage}{0.93\textwidth}
        \centering
        \includegraphics[width=0.96\textwidth]{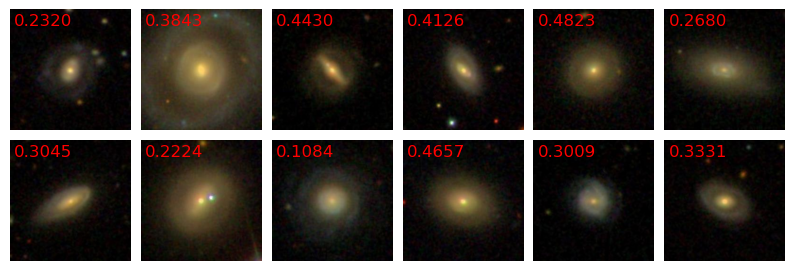}
        False negatives
        \label{fig:fn}
    \end{minipage}

    \caption{Examples of images from our test set that are both correctly and incorrectly classified.}
    \label{fig:val_images}
\end{figure*}

Our trained model obtained a classification accuracy of 98 percent on the testing set. The threshold used for converting prediction scores to class labels is 0.50.
The classification report shown in Figure \ref{fig:train-results} summarizes the precision and recall we have obtained for both the classes. The precision obtained for the rings was 93 percent. Our model also has a good recall value of 89 percent.Using the predicted probabilities of our network, a catalog with higher purity can be obtained by appropriately selecting the classification threshold. It is to be kept in mind that higher purity comes at a cost of reduced completeness.
The nature of the trade-off can be understood using the precision-recall curve shown in Figure \ref{fig:train-results}.
Figure \ref{fig:val_images} shows galaxies in the test set that are correctly classified (True Positives and True Negatives) and failures (False Positives and False Negatives) obtained using our trained network.

\subsection{Prediction at Different Zoom Levels}
\begin{figure*}
    \centering
    \includegraphics[width=0.65\textwidth]{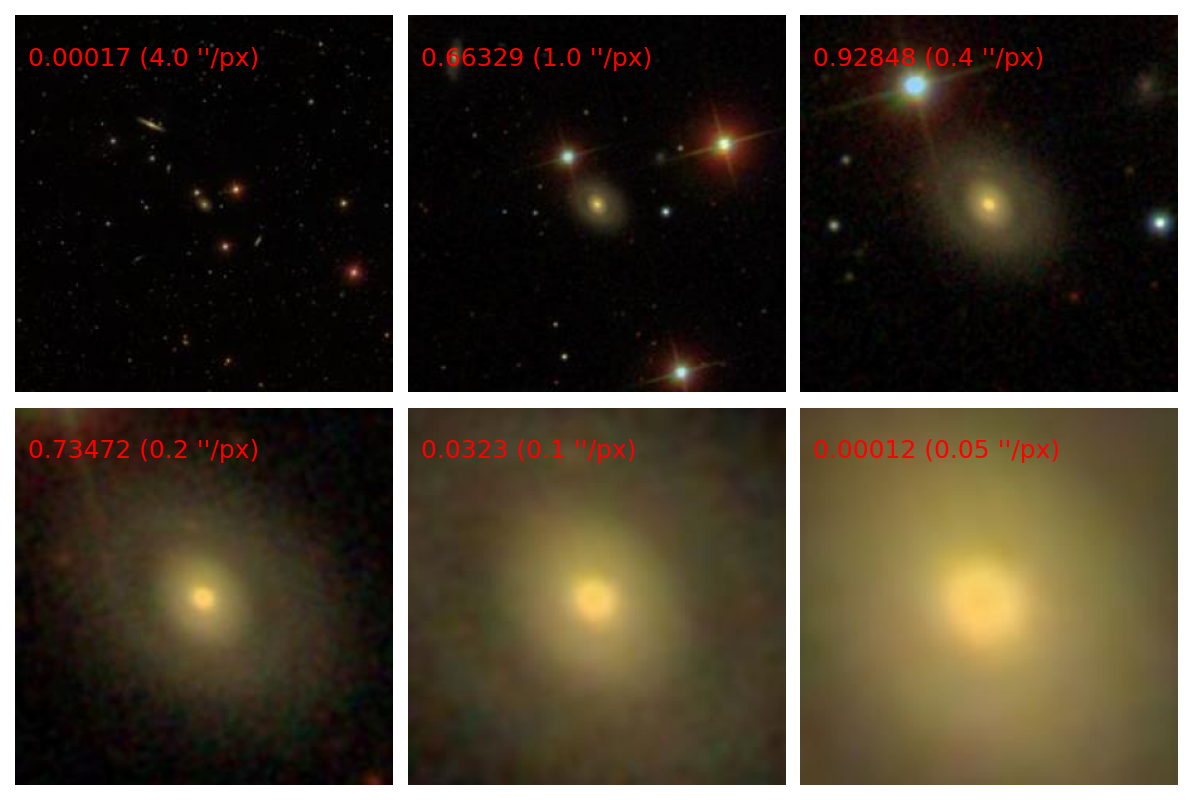}
    \caption{Cutouts of a single galaxy downloaded at various zoom levels from the SDSS cutouts server and predicted using the trained network. The prediction scores obtained for each are shown in red color.}
    \label{fig:zoom_lev}
\end{figure*}

To get a better insight into what the model has learnt we downloaded a single galaxy at different zoom levels using the SDSS image cutout service.
This is done by varying the scale factor and downloading the different images.
The trained network was then used to predict on these images.
Figure \ref{fig:zoom_lev} shows different versions of a particular ringed galaxy along with the probability score returned by the network.
It can be seen that probability score goes to zero both when the images is too zoomed out or zoomed in such that the ring feature is not visible.
For other zoom levels the probability score given by the network are comparable to what a human observer would assign.

\subsection{Analysis of Failed Predictions}
To get a better idea of the  performance of our classifier we analyzed the failed predictions.
A quantitative way of doing this is to compute the log-loss for each sample in our test set.
The log-loss is  defined by 
$$ \rm{logloss_{i}} = -[y_{i} \ln p_{i}+(1-y_{i})\ln (1-p_{i})] $$
where $y_i$ is the true value which is either 0 or 1,  and  $p_i$ is the predicted probability score for observation $i$.
In our binary classification problem, with a sigmoid activation for the last layer, the predicted value is a real valued number between 0 and 1, which is interpreted as a probability of the sample actually belonging to the positive class.
For the negative class a low predicted probability implies a high confidence value.
Hence in this case we  need to take the difference of the value from one in order to get the real confidence.
When analyzing the failures in our testing set we can use a loss function that is weighted using the confidence with which a prediction has been made.

\begin{figure*}
    \begin{minipage}{\textwidth}
    \centering
    \includegraphics[width=\textwidth]{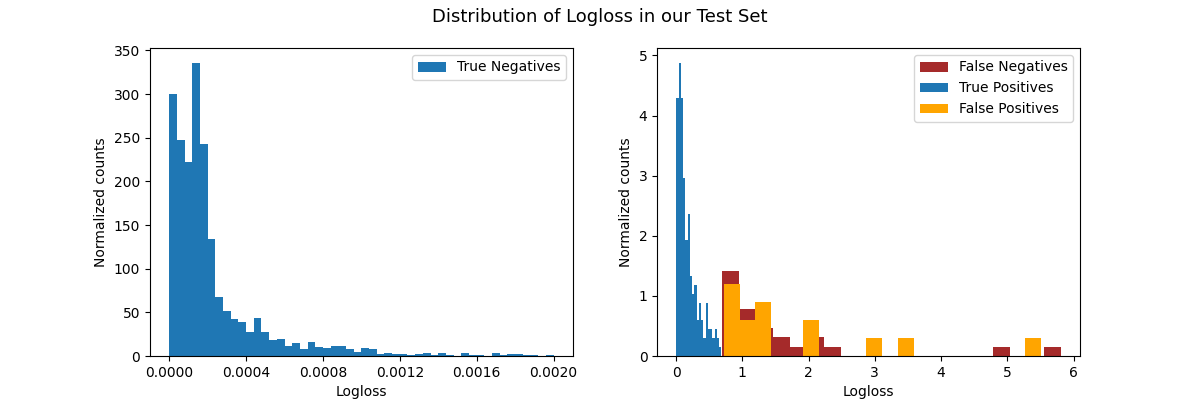}
\end{minipage}
\begin{minipage}{\textwidth}
    \centering
    \includegraphics[width=0.65\textwidth]{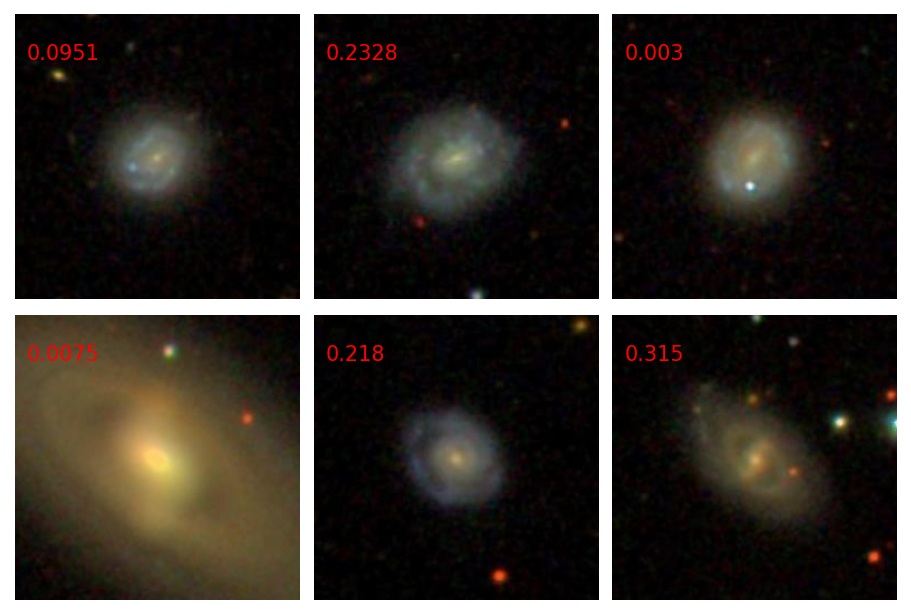}
\end{minipage}
    \caption{The normalized distribution of log-loss in our test set (above). The true negatives are shown separately because of the difference in bin range. A sample of galaxies from the false negatives having the highest log-loss values (below). The annotations shown in red color denote the prediction scores.}
    \label{fig:test_hist_fn}
\end{figure*}

The histogram of these losses can then be used to get an idea about the overall performance of the model.
Figure \ref{fig:test_hist_fn} shows the histogram of log-loss obtained for our test set.
The peak of the histograms are towards the lower end,
which means that there are a lesser number of wrong predictions made with a high confidence value and greater number of correct predictions with a high confidence value.
That is, it can be seen that the confidence of wrong prediction is much lower than that of the correct predictions.
The figure also shows representative images from our test set corresponding to the false negatives and having high log-loss values.
The rings in these galaxies are not very distinct and one of the galaxies has a high inclination angle that might have confused the network.

\subsection{Model Interpretability}
\begin{figure*}
    \begin{minipage}{\textwidth}
    \centering
        \includegraphics[width=\textwidth]{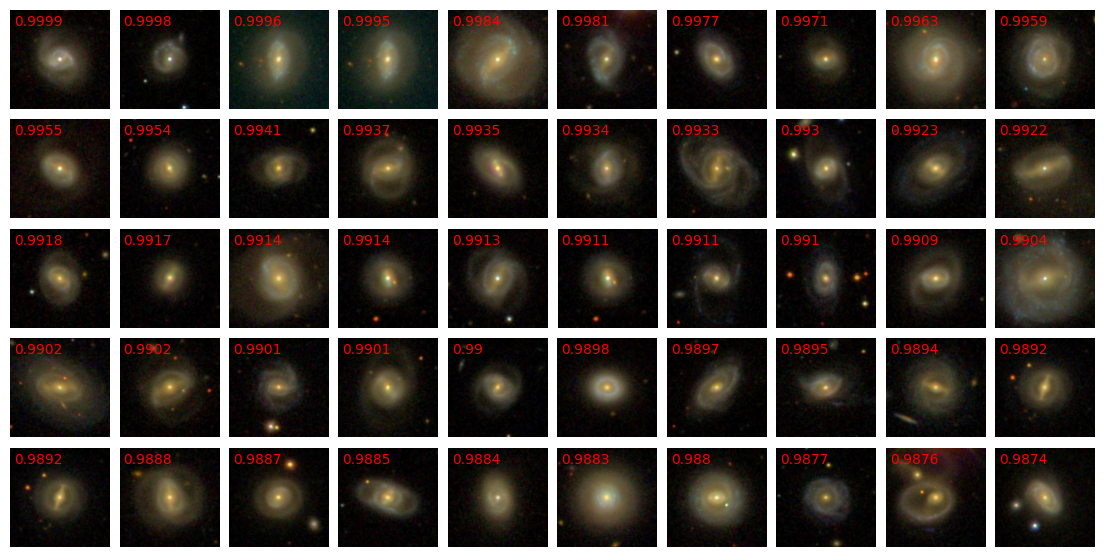}
	(a)
        \label{fig:train_lowloss_new}
    \end{minipage}
    \begin{minipage}{\textwidth}
    \centering
        \includegraphics[width=\textwidth]{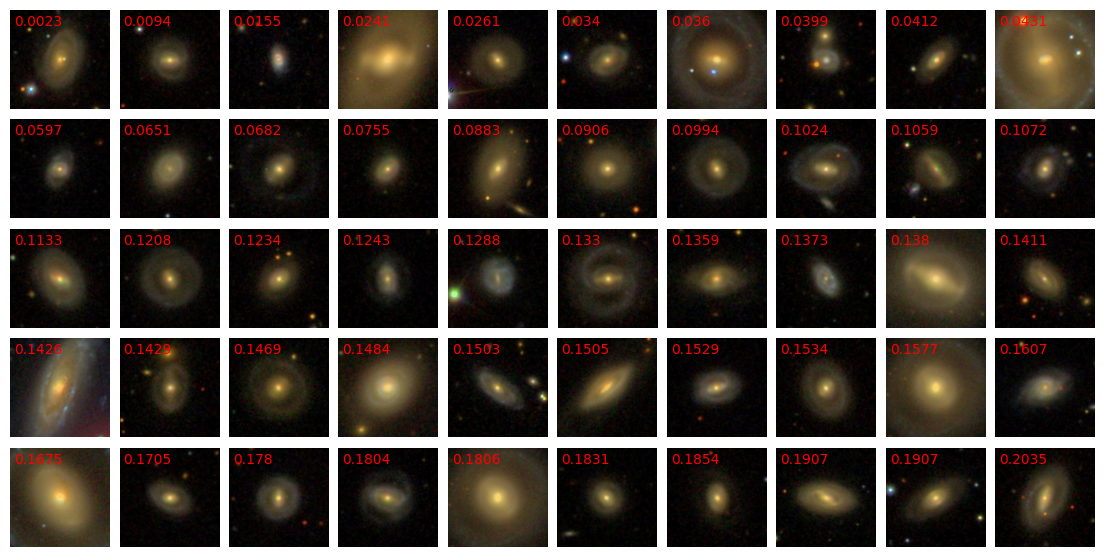}
	(b)
        \label{fig:train_highloss_new}
    \end{minipage}
    \caption{Ringed galaxies from our training sample predicted using the trained model and with a low log-loss (above) and high log-loss (below). The annotations shown in red colour denote the prediction scores.
}    \label{fig:train_high_low_loss}
\end{figure*}

Once the model is trained, if we use it to make predictions on the training data itself, we will be able to get some idea of what the model has learnt, which is known as model interpretability.
We used the trained model to predict the probability of each ringed galaxy in the sample  being identified as ringed. The difference (1-predicted probability) can be taken as a measure of the error.   
This error can be used for finding subsets of ringed galaxies in our training sample that are either easy for the network to learn or difficult to learn.
The results for a sample are shown in Figure \ref{fig:train_high_low_loss}.
It can be seen that among the galaxy images that the network had trouble in learning, there are many with image artefacts or the presence of background stars or galaxies.
Many of the galaxies have a zoomed in view as well.
The galaxy images that were easily identified by the network had a better contrast with the background and the galaxies were at a normal zoom level.

\section{A Catalog of Ringed Galaxies} \label{catalogue_rings}

\begin{figure*}
    \begin{minipage}{\textwidth}
    \centering
        \includegraphics[width=0.95\textwidth]{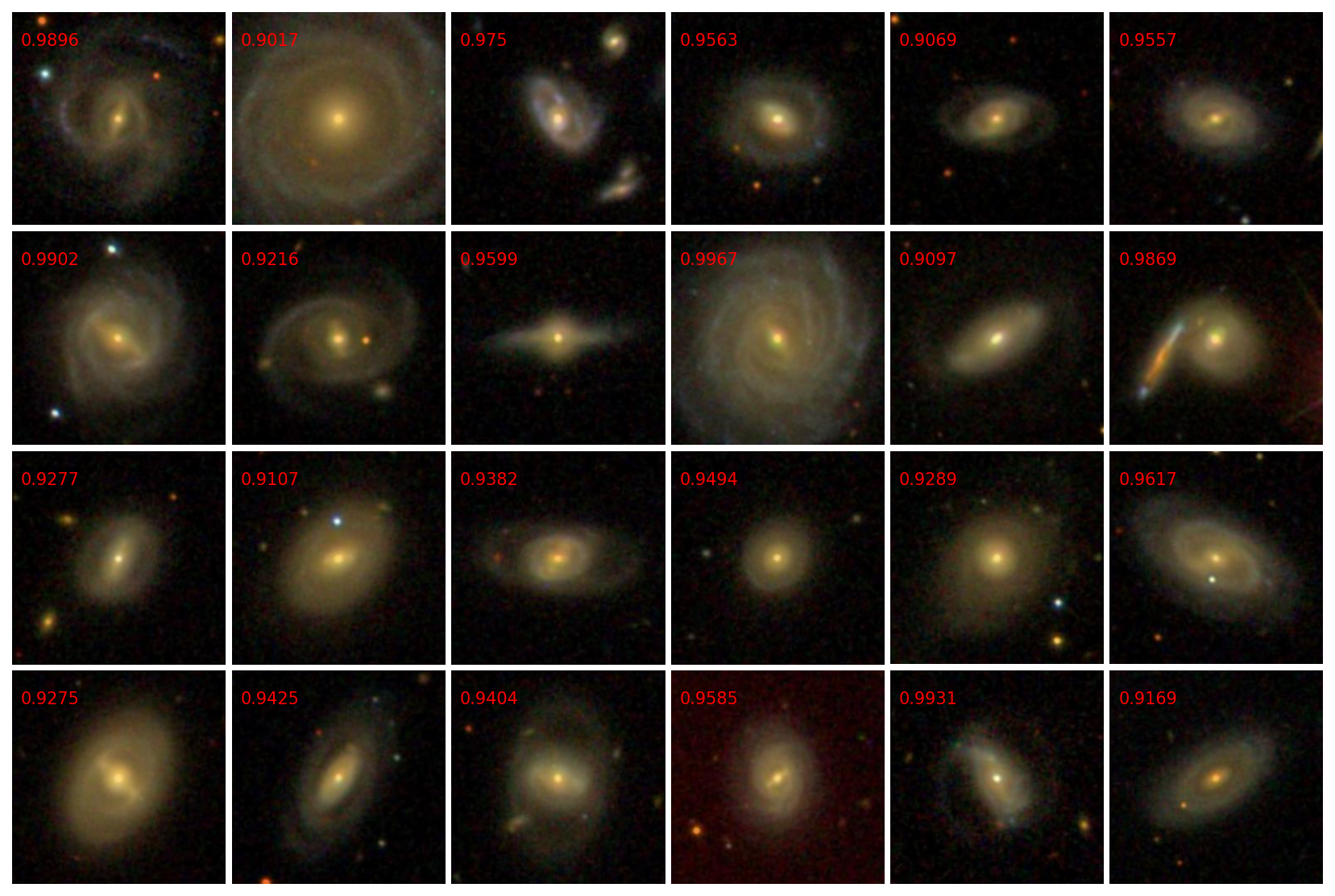}\\
	(a)
        \label{fig:catalog_rings}
    \end{minipage}
    \begin{minipage}{\textwidth}
    \centering
        \includegraphics[width=0.95\textwidth]{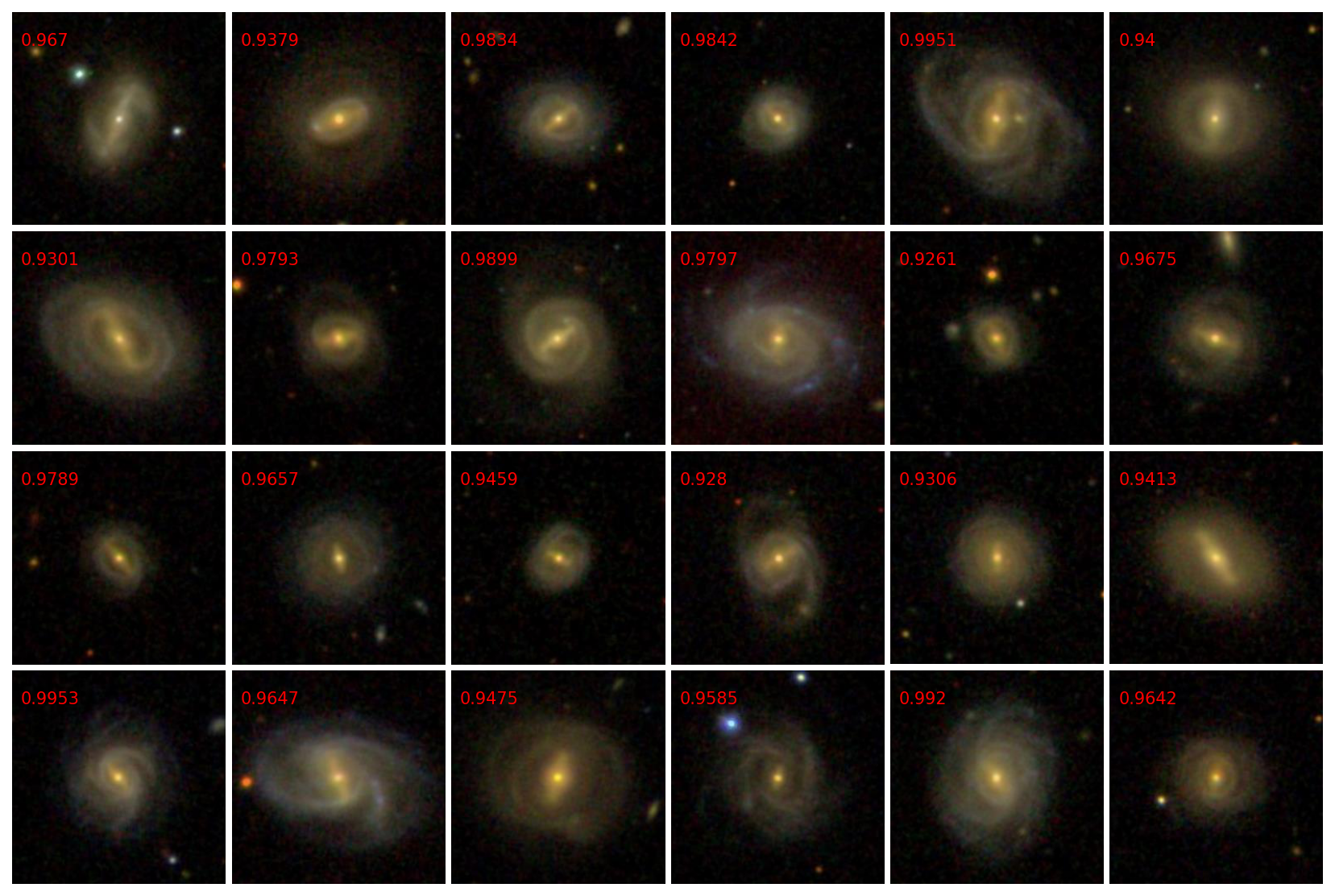}\\
	(b)
        \label{fig:catalog_barred_rings}
    \end{minipage}
    \caption{Some good examples of ringed galaxies (above) and barred ring galaxies(below) from our catalog. The annotations shown in red colour denote the prediction scores. }
    \label{fig:catalog_images}

\end{figure*}

To prepare a catalog of ringed galaxies from SDSS we used the same selection criteria that was used for the training data preparation.
The SDSS data can be queried using the CASJOBS \footnote{\url{https://skyserver.sdss.org/casjobs/default.aspx}} interface.
The complete query is given in the Appendix \ref{query}.
The source list obtained from the CASJOBS server was downloaded using our bulk download script.
The trained network was then used to predict the ringed or non-ringed nature of these galaxies.
A preview of the catalog is included in Table \ref{fig:two-cats}.
Figure \ref{fig:catalog_images} shows a selection of galaxies with rings taken from our catalog.

\begin{figure*}
    \centering
    \includegraphics[width=\textwidth]{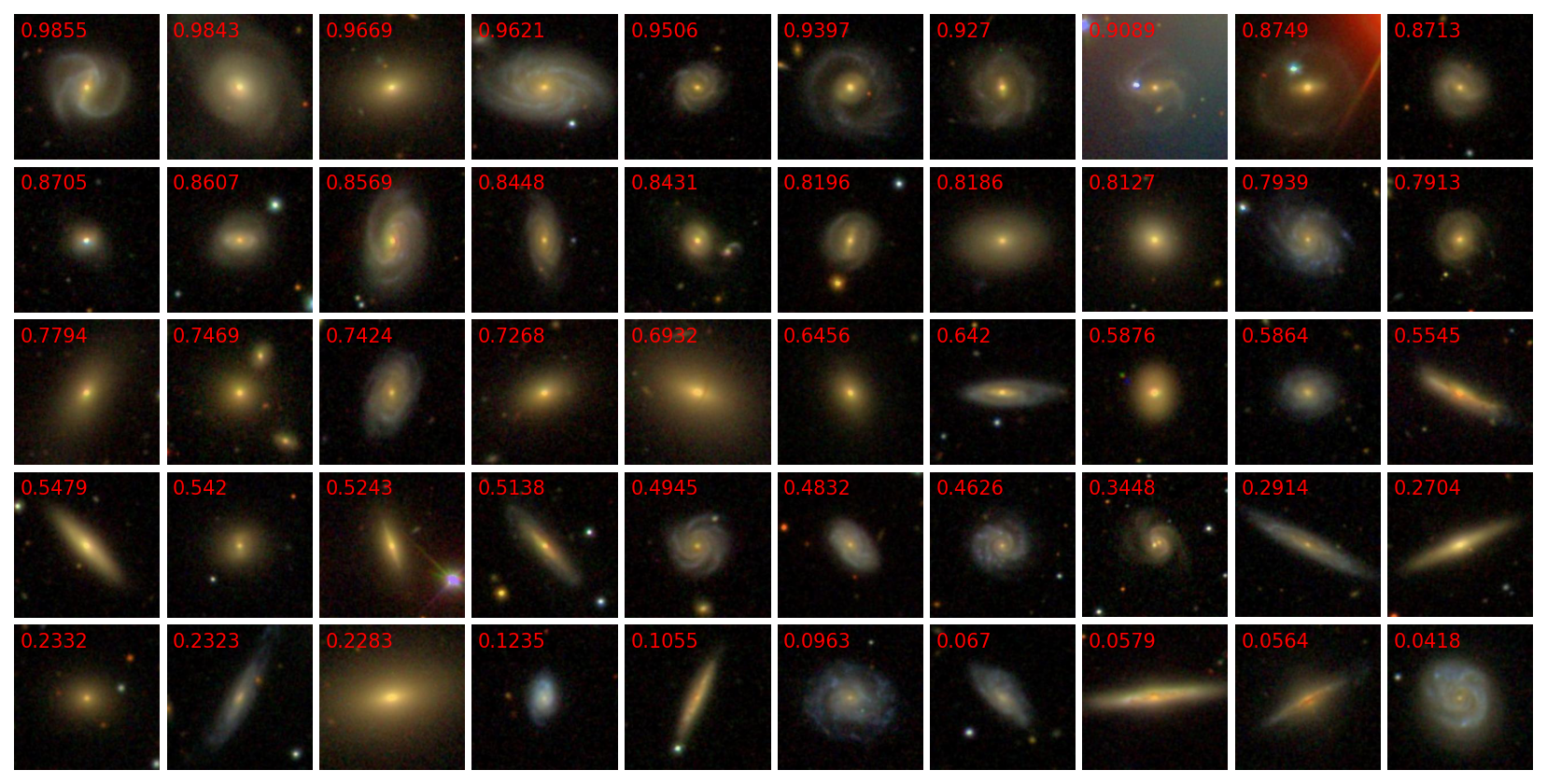}
    \caption{A mosaic of 50 randomly selected images from the prediction set sorted using the predicted probability of having a ring. Visual inspection shows that the rings have accumulated towards the bottom and the non-rings towards the top.}
    \label{fig:pred_mosaic}
\end{figure*}

\begin{figure}
    \includegraphics[width=\linewidth]{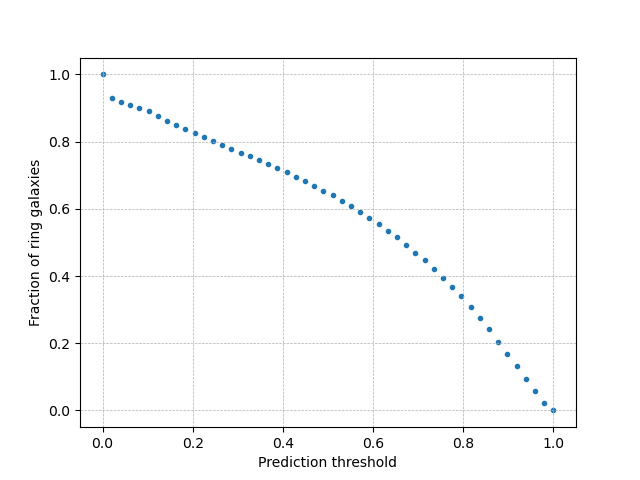}
    \caption{Plot showing the fraction of ring galaxies predicted at various thresholds.}
    \label{fig:frac_ring_gal}
\end{figure}

\begin{figure*}
    \includegraphics[width=0.5\linewidth]{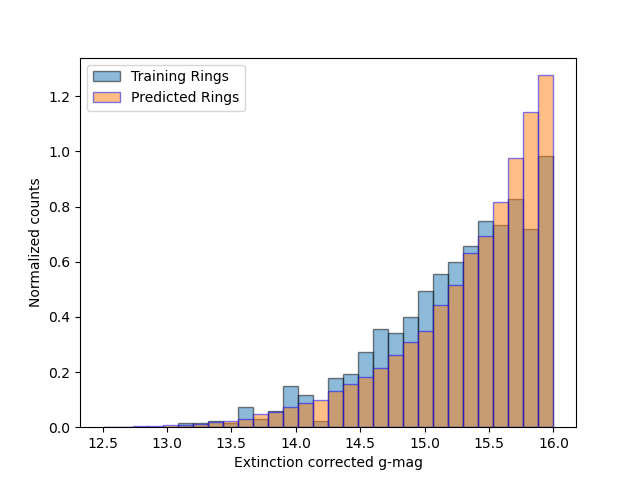}
    \includegraphics[width=0.5\linewidth]{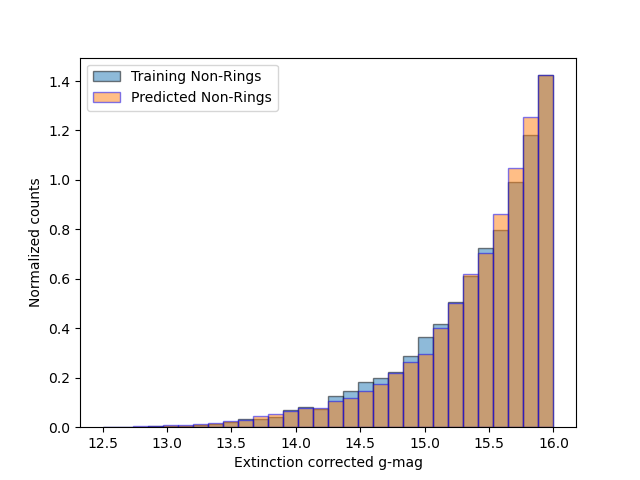}
    \caption{A comparison of the extinction corrected g-magnitude distribution between the training and the prediction samples shown for both the ringed galaxies (left) and galaxies without rings (right). }
    \label{fig:gmag_dist_train_pred}
\end{figure*}

In order to do a more thorough testing of our network on real world data, we show in Figure \ref{fig:pred_mosaic} an image mosaic created from 50 randomly selected galaxies from our prediction set along with the
machine generated probability scores.
The galaxies are  sorted in descending order of the prediction score.
Based on the natural abundance of ringed galaxies, we may expect less than 10 of these to be ringed.
It can be seen on visual inspection that most of the galaxies with true rings have a high prediction scores in comparison to the non-rings and hence they are clustered towards the top of the image mosaic.
We found 4855 galaxies that are predicted to have rings using a classification threshold of 0.90.
This corresponds to a ring galaxy fraction of 16.5 percent. At a slightly lower threshold of 0.89 we recover the same ring galaxy fraction of 18.2 as observed by \citet{nair_catalog_2010}.
Figure \ref{fig:frac_ring_gal} shows how this fraction varies for all thresholds between 0 and 1.
Further in Figure \ref{fig:gmag_dist_train_pred} we show how the distribution of the extinction corrected g-magnitude of the predictions compares with that of the training sample.

\begin{table*}[htbp]
    \centering
        \begin{tabular}{rrrrrrrrl}
\multicolumn{9}{r}{} \\
\toprule
ra & dec & objid & gmag & deVRad\_r & deVAB\_g & redshift & Prediction & Label \\
\midrule
40.285690 & -0.714957 & 1237645941824356443 & 15.641980 & 9.327478 & 0.339572 & 0.040287 & 0.908651 & Rings \\
57.025337 & 0.208845 & 1237645942905438473 & 15.497970 & 10.815760 & 0.871453 & 0.025475 & 0.598131 & NonRings \\
56.781387 & 1.000343 & 1237645943979114582 & 15.879010 & 2.804902 & 0.321389 & 0.039847 & 0.250911 & NonRings \\
56.847420 & 0.875488 & 1237645943979114622 & 15.182350 & 13.953940 & 0.724238 & 0.039371 & 0.865321 & NonRings \\
57.248385 & 0.925979 & 1237645943979311221 & 15.657840 & 6.806947 & 0.535690 & 0.035788 & 0.818097 & NonRings \\
57.674720 & 1.040755 & 1237645943979507954 & 15.344130 & 3.541592 & 0.822157 & 0.037123 & 0.516564 & NonRings \\
243.708876 & -0.915653 & 1237648672921485632 & 15.293520 & 8.426456 & 0.640077 & 0.030767 & 0.836999 & NonRings \\
246.015172 & -0.902869 & 1237648672922468973 & 15.979040 & 15.339940 & 0.550030 & 0.046526 & 0.494671 & NonRings \\
245.367353 & -0.457074 & 1237648673459077169 & 15.901310 & 8.323227 & 0.500005 & 0.059092 & 0.534696 & NonRings \\
246.782081 & -0.492432 & 1237648673459667234 & 15.630960 & 8.579305 & 0.437323 & 0.046192 & 0.457613 & NonRings \\
189.522249 & -0.027031 & 1237648673971437623 & 14.639720 & 18.461840 & 0.442339 & 0.012513 & 0.802063 & NonRings \\
237.945144 & -0.105170 & 1237648673992671592 & 15.529860 & 5.215971 & 0.949471 & 0.054426 & 0.630027 & NonRings \\
243.236782 & -0.096251 & 1237648673994965546 & 14.331720 & 27.210810 & 0.723929 & 0.030867 & 0.879053 & NonRings \\
243.583196 & -0.031564 & 1237648673995162093 & 15.782320 & 5.114430 & 0.559130 & 0.030811 & 0.536166 & NonRings \\
245.381888 & -0.072364 & 1237648673995948107 & 15.426320 & 3.416426 & 0.888818 & 0.027287 & 0.904141 & Rings \\
248.064172 & -0.049932 & 1237648673997127724 & 15.503360 & 5.104637 & 0.811834 & 0.044017 & 0.924116 & Rings \\
195.644361 & 0.348565 & 1237648674510995594 & 14.989240 & 7.022532 & 0.659088 & 0.067754 & 0.922850 & Rings \\
242.672630 & 0.276520 & 1237648674531639653 & 15.742830 & 5.721440 & 0.605837 & 0.062059 & 0.449378 & NonRings \\
243.154609 & 0.379972 & 1237648674531836029 & 15.808500 & 3.519223 & 0.956060 & 0.043145 & 0.945323 & Rings \\
249.820694 & 0.410283 & 1237648674534719840 & 15.602230 & 4.794272 & 0.273508 & 0.024328 & 0.069281 & NonRings \\
\bottomrule
\end{tabular}

    \vspace{1cm}
        \begin{tabular}{rrrrl}
\multicolumn{5}{r}{} \\
\toprule
ra & dec & SDSS\_Objid & Prediction & Label \\
\midrule
0.013400 & -1.113000 & 1237663275780276407 & 0.001809 & Barred-NonRings \\
0.019800 & 0.781700 & 1237657191978959126 & 0.650422 & Barred-Rings \\
0.032300 & -0.723700 & 1237663783123681369 & 0.039325 & Barred-NonRings \\
0.056000 & -1.213600 & 1237663275780276438 & 0.060768 & Barred-NonRings \\
0.114500 & 14.962500 & 1237656495650570466 & 0.208050 & Barred-NonRings \\
0.117000 & 14.381200 & 1237652942638481637 & 0.183115 & Barred-NonRings \\
0.128700 & -1.213000 & 1237663275780341888 & 0.532812 & Barred-Rings \\
0.158900 & 14.623600 & 1237656495113699423 & 0.024679 & Barred-NonRings \\
0.166500 & -1.191200 & 1237663275780341936 & 0.036194 & Barred-NonRings \\
0.228100 & 15.218700 & 1237652943712288939 & 0.000647 & Barred-NonRings \\
0.241000 & 14.190400 & 1237656494576894171 & 0.162884 & Barred-NonRings \\
0.241700 & 14.864400 & 1237652943175418025 & 0.215693 & Barred-NonRings \\
0.261300 & -0.257100 & 1237663783660617841 & 0.001189 & Barred-NonRings \\
0.297800 & 15.985900 & 1237656496724443439 & 0.000518 & Barred-NonRings \\
0.300100 & 15.351400 & 1237652943712288996 & 0.173043 & Barred-NonRings \\
0.306000 & -0.992900 & 1237657189831606381 & 0.008202 & Barred-NonRings \\
0.368800 & -0.224400 & 1237663783660683434 & 0.321357 & Barred-NonRings \\
0.380900 & 14.407500 & 1237652942638612625 & 0.211441 & Barred-NonRings \\
0.388600 & -0.730500 & 1237663783123812523 & 0.000156 & Barred-NonRings \\
0.398900 & -0.395200 & 1237663783660683462 & 0.303564 & Barred-NonRings \\
\bottomrule
\end{tabular}

\caption{A preview of the catalog of ringed galaxies (above) and of barred rings (below) with the first 20 rows from each catalog shown.}
    \label{fig:two-cats}
\end{table*}

    \section{A Catalog of Ringed galaxies with Bars} \label{catalogue_barred_rings}
    Bars in galaxies are much more common than rings. Approximately half of the observed galaxies are seen to have bars \citep{knapen_observations_1999}. However galaxies that contain both bars and rings are certainly a smaller fraction. Dedicated catalogs for such galaxies are also quite rare.
    Here we use the \cite{abraham_detection_2018} catalog, which is an automated catalog of galaxies with bars in SDSS, to find galaxies with both bars and rings.
    The \cite{abraham_detection_2018} catalog contains 111,838 galaxies that have been predicted as either barred or unbarred. 
    We selected the barred galaxies which came to be 25,781 in number and used our trained network to predict which of these galaxies also have rings.
    Figure \ref{fig:catalog_images} shows a random selection of the barred galaxies that have been predicted to have rings with a probability score of greater than 90 percent.
    We found 2087 galaxies that contained both bars and rings using a classification threshold of $0.5$.
    A preview of the catalog containing barred galaxies with predictions for the presence or absence of a ring is shown in  Table \ref{fig:two-cats}.

\section{Discussion} \label{discussion}
As a demonstration of the usefulness of the catalogs generated in this paper, in this section, we will explore the environments and star formation characteristics of ring galaxies in our sample. Our goal is to analyze star formation within the framework of the main sequence of star-forming galaxies, the transitioning population of green valley galaxies, and the quenched galaxies. Additionally, we will examine star formation in relation to the environmental variations from low to high densities. For this analysis, we use a sample of 4855 galaxies with a classification threshold of 0.90 that is predicted to have rings.  

We cross-matched this sample with \citet{2006MNRAS.373..469B} to find local surface density as a measure of the environment. It is calculated using the relation $\Sigma_\text{N} = \text{N}/\pi d_\text{N}^2$, where $d_\text{N}$ represents the distance to the Nth nearest neighbour. These neighbours fall within the redshift range $\pm \Delta zc = 1000\ \text{km/s}$ for galaxies with spectroscopic redshifts or within the 95\% confidence interval for galaxies with only photometric redshifts. We use the best estimate, $\Sigma$, which is the average of $\Sigma_N$ for the 4th and 5th nearest neighbours and categorize our sample into low-density ($\log \Sigma \ (\text{Mpc}^{-2}) < -0.5$), intermediate-density (-0.5 < $\log \Sigma \ (\text{Mpc}^{-2}) < 0.5$), and high-density ($\log \Sigma \ (\text{Mpc}^{-2}) > 0.5$) groups. 

\begin{figure}
\includegraphics[width=\linewidth]{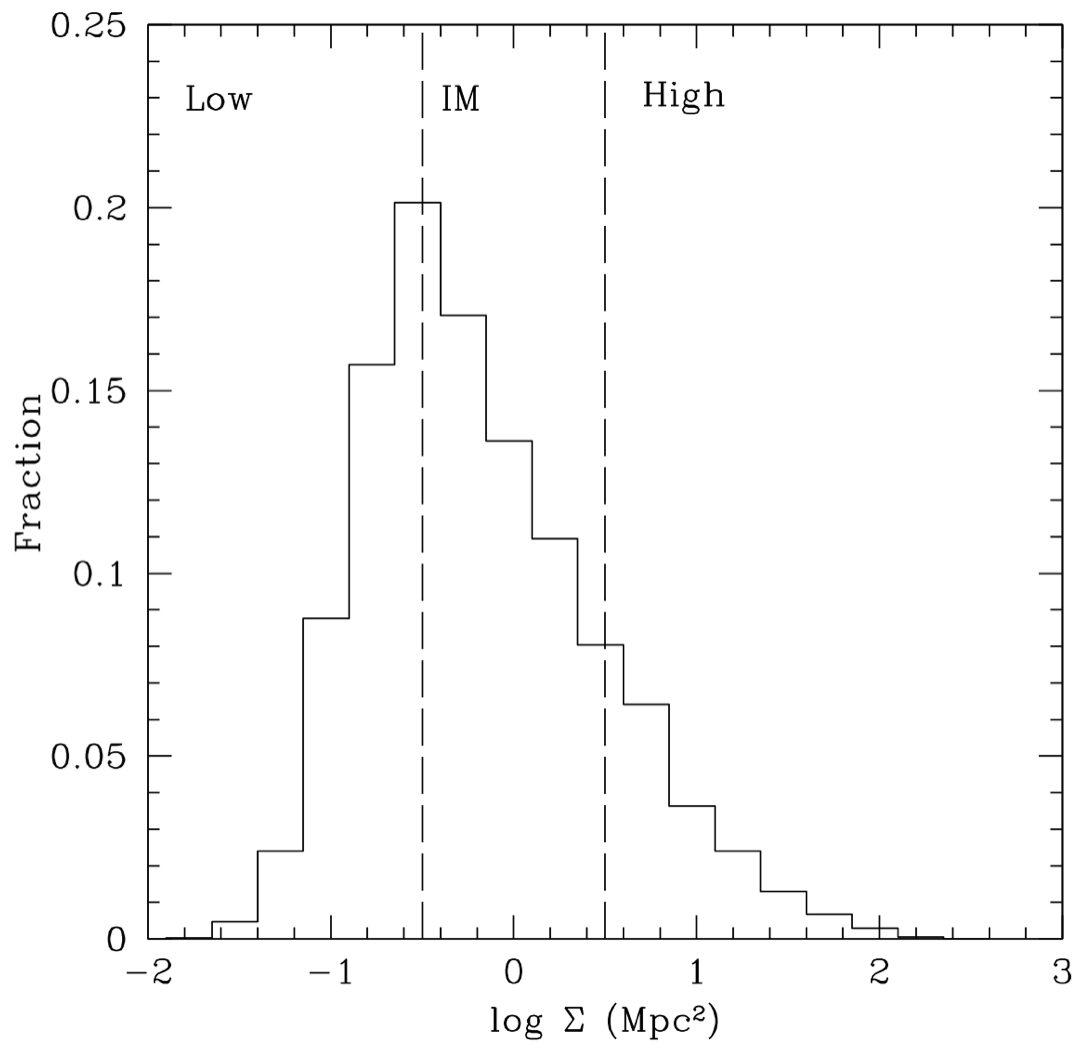}
\caption{Histogram of local environmental density ($\log \ \Sigma \ (\text{Mpc}^{-2})$) for our final sample of galaxies. The dashed lines separate the sample into low ($\log \Sigma \ (\text{Mpc}^{-2}) < -0.5$), intermediate(-0.5 < $\log \Sigma \ (\text{Mpc}^{-2}) < 0.5$), and high densities($\log \Sigma \ (\text{Mpc}^{-2}) > 0.5$).} 
\label{fig:Figure_Ring-Env}
\end{figure}

Figure \ref{fig:Figure_Ring-Env} presents the histogram of $\log \ \Sigma$ for the sample of 4471 ring galaxies with environment information. We find that ring galaxies are primarily found in intermediate density environments, i.e. in galaxy groups  (49.3\%) and in low-density environments, i.e. as isolated galaxies 35.1\%). We see relatively few ring galaxies (15.6\%) in high-density environments in agreement with \citet{1992A&A...257...17E}. Numerical simulations, such as ROMULUS \citep{2017MNRAS.470.1121T}, suggest that galaxies within group environments often undergo interactions or multiple mergers with gas-rich satellites, evolving into elliptical galaxies. Subsequently, these galaxies start to accrete gas and redevelop their disks, which then become unstable, triggering star formation in ring-like patterns. When galaxies display rejuvenated star-forming rings, two pathways for gas inflow are identified in numerical simulations: (a) Diffuse cooling gas that feeds the galaxy, aiding in developing the disk/ring structure. (b)The acquisition of ram pressure-stripped gas streams from gas-rich satellites \citep{2022MNRAS.515...22J}. 

To study star formation in our sample ring galaxies with environment information, we utilized the GALEX-SDSS-WISE Legacy Catalog (GSWLC-X2) \citep{2016ApJS..227....2S,2018ApJ...859...11S} and the cross-match gave us 4022 ring galaxies with a stellar mass (log M$_{*}$) range of 8.8-11.68. To investigate the influence of the environment on the star formation process in ring galaxies, we constructed a control sample of non-ring galaxies with a classification threshold of 0.40 for being a ring galaxy, i.e., a threshold of 0.60 for being a non-ring galaxy. This control sample has a similar r-band magnitude and redshift, and a cross-match with the GSWLC-X2 catalog gave $\sim$5000 non-ring galaxies. The GSWLC-X2 catalog integrates ultraviolet data from GALEX, optical data from SDSS, and infrared data from the WISE all-sky survey to model galaxy spectral energy distribution (SED). During SED modelling, the catalog applies flexible dust attenuation laws and emission line corrections to ensure accurate measurements of star formation rate (SFR) and stellar mass (M$_{*}$), as it might lead to the wrong identification of galaxies.  

\begin{figure*}
\includegraphics[width=\textwidth]{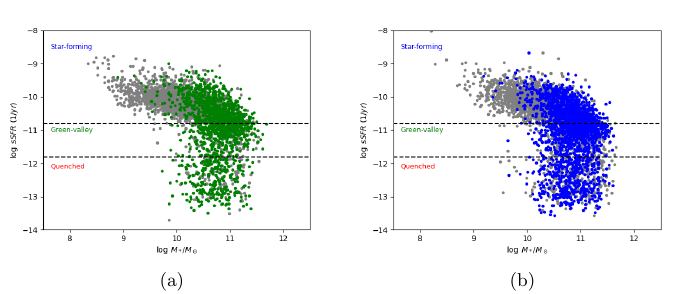}
    \centering
\includegraphics[width=0.5\textwidth]{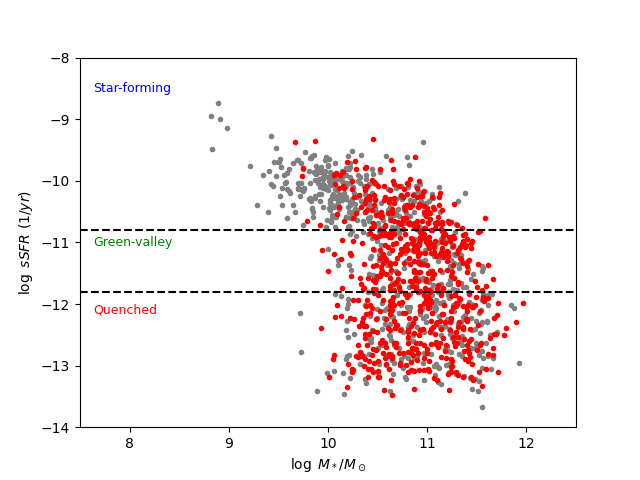}\\
{\Large (c)}
\caption{Dependence of specific star formation rate on stellar mass (sSFR-Mass plane) for our sample of ring galaxies and control sample of non-ring galaxies (gray dot) as a function of (a)  low-density ($\log \Sigma \ (\text{Mpc}^{-2}) < -0.5$) (b)  intermediate-density (-0.5 < $\log \Sigma \ (\text{Mpc}^{-2}) < 0.5$), and (c) high-density environment ($\log \Sigma \ (\text{Mpc}^{-2}) > 0.5$)}
\label{fig:Figure_ring-sSFR}
\end{figure*}

We calculated the specific star formation rate (sSFR) and, following \citet{2014SerAJ.189....1S}, classified our sample galaxies into three sub-classes  as a function of environment, as shown in Figure \ref{fig:Figure_ring-sSFR}. The star-forming region is identified with log sSFR $\geq$ -10.8, the green valley, which is a transition zone between the star-forming and quenched regions, is defined as -11.8 < log sSFR < -10.8, and quenched region is defined by log sSFR $\leq$ -11.8. In high-density environments, most of the ring galaxies are in the quenched (42.3 \%) or green valley region (33.6 \%); however, the control sample of non-ring galaxies are almost equally distributed in quenched (33.8\%) and star-forming region (47.8\%) with fewer galaxies in green valley (18.9\%). In contrast, low and intermediate-density environments are dominated by star-forming ring galaxies (52.4\% and 43.2 \%, respectively), as the non-ring galaxies from the control sample (86.9\% and 76.3 \%, respectively). However, the number of ring galaxies in green valleys increases for intermediate-density environments, though the number of non-ring galaxies increases for high-density environments.   It is apparent that most of the ring galaxies are located in the star-forming region for low and intermediate-density environments, which validates the scenario in which rejuvenation of star formation happens by accreting gas and triggering star formation in ring-like features. Our results are somewhat diﬀerent from those of \citet{2024A&A...683A..32F}, who suggest that ringed galaxies exhibit lower star formation activity in intermediate-density environments, though it is much lower in high-density environments.

Many ring galaxies from our sample are also part of the MaNGA survey, and the study of spatially resolved stellar populations of rings using such  IFU surveys will provide useful insight into the origin of star-forming rings, and this can be the subject of future work.


\section{Conclusion} \label{conclusion}

We have shown that a relatively simple deep neural network based on the AlexNet architecture can be used to detect galactic rings from colour composite galaxy images.
The training required only a modest sample size of 1122 original ringed galaxy images and about 10639 original images of galaxies without rings.
To overcome the challenges of training with a small dataset of labelled rings, we used a smaller network and also increased the total number of images available for training through realistic image transformations.
We also used evaluation metrics suited for the class imbalance that was present in our data.
We have prepared a catalog of candidate ring and non-ring galaxies using the predictions of our trained network.
The probability score provided in the catalog can be used as a confidence measure of the presence of rings, to obtain smaller catalogs with the required level of confidence for follow up analysis.
Additionally we used the network to generate a catalog of galaxies with rings as well a bars, which may be of special interest to galaxy morphologists. Finally, we explored the connection between the environments and star formation characteristics of the 4855 ringed galaxies in our catalog. We found that ring galaxies are star-forming and primarily found in low- and intermediate-density environments.

\appendix
\section{CASJOBS query}
\label{query}
We provide the CASJOBS query used to create the catalog of ringed and non-ringed galaxies.
The query is run in the DR18 context.
\begin{verbatim}
SELECT
  p.objid,p.ra,p.dec,
  (p.u-p.extinction_u) as umag,
  (p.g - p.extinction_g) as gmag,
  (p.r-p.extinction_r) as rmag,
  (p.i-p.extinction_i) as imag,
  (p.z - p.extinction_z) as zmag,
  (p.deVRad_g) as g_dev, p.deVRad_r, p.deVAB_g,
  s.specobjid, s.class,
  s.z as redshift into mydb.MyTable
  from PhotoObjAll AS p
JOIN SpecObj AS s ON s.bestobjid = p.objid
WHERE 
  s.class='GALAXY' 
  AND s.z BETWEEN 0.01 and 0.1
  AND (p.g - extinction_g) < 16

\end{verbatim}

\bibliography{zotero}
\bibliographystyle{aasjournal}
\label{lastpage}
\end{document}